\documentclass[sigconf,authorversion,nonacm]{acmart}
\usepackage{array}
\usepackage{multirow}
\usepackage{subcaption}

\PassOptionsToPackage{table,xcdraw}{xcolor}

\AtBeginDocument{%
  \providecommand\BibTeX{{
    \normalfont B\kern-0.5em{\scshape i\kern-0.25em b}\kern-0.8em\TeX}}}

\setcopyright{none}
\acmYear{2023}

\acmConference[TAICHI '23]{TAICHI '23: Taiwanese Association of Computer Human Interaction}{August 19--20, 2023}{Taipei City}
\renewcommand\footnotetextcopyrightpermission[1]{} 

\begin{document}

\newcommand{\projectName}{BetterMinton Service}

\title[\projectName{}]{\projectName{}: Analyzing the Badminton Service using Open Kinetic Chain}

\author{Eden Cong-He Xu}
\affiliation{
  \institution{National Taiwan University}
  \department{Department of Computer Science and Information Engineering}
  \streetaddress{No. 1, Sec. 4, Roosevelt Rd.}
  \city{Taipei}
  \country{Taiwan}
  \postcode{10617}
}
\email{edencxu@gmail.com}

\author{Lung-Pan Cheng}
\orcid{0000-0002-7712-8622}
\affiliation{
  \institution{National Taiwan University}
  \department{Department of Computer Science and Information Engineering}
  \streetaddress{No. 1, Sec. 4, Roosevelt Rd.}
  \city{Taipei}
  \country{Taiwan}
  \postcode{10617}
}
\email{lung-pan.cheng@hci.csie.ntu.edu.tw}

\renewcommand{\shortauthors}{Xu and Cheng}

\begin{abstract}
    We present a badminton training system that focuses on the backhand short service.
    Unlike the prior motor skill training systems which focus on the trainee's posture, our system analyzes the process of moving joints with the open kinetic chain (OKC), which helps align movement and minimize muscle use for better joint control.
    We process the users’ mocap data to visually show their last service process comparing to 4 ideal OKC characteristics that we collected from a 6-sub-elite formative study as well as recommended contact posture.
    We validate our system through a 12-user study that measures serving accuracy, qualitative feedback, and skeletal data with users at various skill levels and open source our skeletal analysis model for future use.
    While the participants' overall service accuracy was not significantly improved, our results show that our system helps participants in the short term to fine-tune their service motion closer to our ideal 4 OKC characteristics. 
\end{abstract}

\ccsdesc[500]{Applied computing~Computer-assisted instruction}
\ccsdesc[500]{Human-centered computing~Interactive systems and tools}
\ccsdesc[500]{Human-centered computing~Empirical studies in HCI}
\ccsdesc[500]{Human-centered computing~Empirical studies in interaction design}

\keywords{badminton backhand short service, motor skill training, open kinetic chain}

\begin{teaserfigure}
  \includegraphics[width=\textwidth]{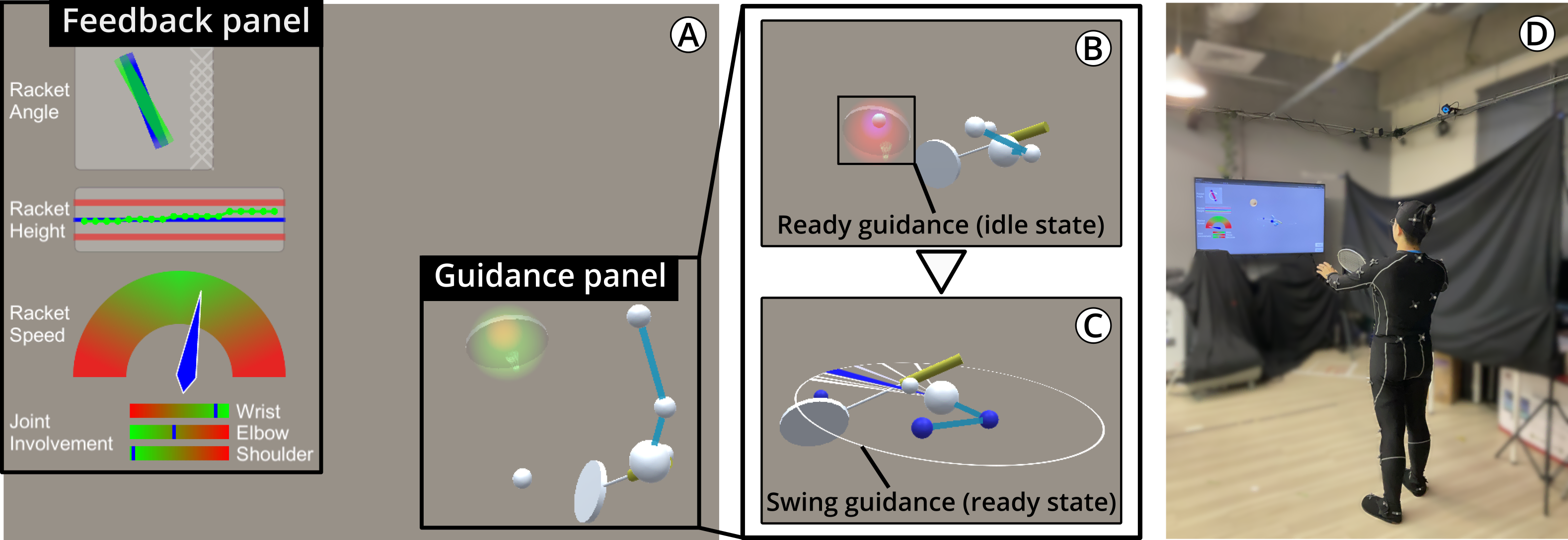}
    \caption{
    We introduce a badminton backhand short service training system focusing on fine-tuning continuous motions for novices.
    This system is built with our analysis of 6 sub-elite players' open kinetic chain characteristics and kinetic variables, such as (1) the racket angle, (2) the racket height, (3) the racket speed, and (4) the involvement of the moving joints.
    We design guidance and feedback to visualize these features so the user can follow and improve their motion.
    (A) is the training system interface displaying the service's guidance and feedback.
    In addition, the user's and the racket's motions that are bound to indicators are illustrated in the guidance panel.
    (B) depicts the ready guidance that tells the racket position is off from the ideal position.
    After the user holds the shuttlecock and the racket at the suggested position, the view turns into the swing guidance illustrated in (C).
    (D) is a user practicing service with the system.
    }
  \Description{The user interface of the system. A user is practicing with BetterMinton Service.}
  \label{fig:figure1}
\end{teaserfigure}

\maketitle

\section{Introduction}


Badminton is one of the most popular sports, with approximately 220 million people engaging in it around the world~\cite{phomsoupha2015science, bwf:wilkieanalysis}.
As more people go in for badminton, coaching resources are becoming scarce in quality and quantity, especially for the novice~\cite{he2014development}.
Without guidance and instant feedback from the coach, the beginner may get used to improper motions and needs to take the risk of performing wrong movements that may hurt them in the worst case~\cite{tang2015physio, seroyer2010kinetic, chen2013computer-assisted}.

Computer-aided coaching systems have been proposed to solve the scarcity of coaching resources.
For instance, researchers have introduced a swing training system~\cite{raina2017combat} and an agility training system~\cite{sung2017implementation} in coaching badminton.
Nevertheless, the motions in badminton are considerably diverse.
A player can play a stroke from any corner, side, or middle of the court to any position on the opposing court in forehand or backhand, overhead or underhand.
The variety of motions in badminton makes it challenging to include all these strokes in a single coaching system.
Therefore, most prior works still require a trainer or an expert to perform a correct motion for the trainee to mimic before the training to coach general badminton motion.

Among all the strokes, service is one of the most crucial strokes in badminton, and its importance has been more and more emphasized since Badminton World Federation (BWF) adopted a new scoring system in 2006. 
The new scoring system states that the winner of a rally scores a point where a rally, a series of shots between opposing players, must start with the service and is not allowed to have a second service. 
That is, a badminton player needs to launch each shuttlecock accurately.
There are four main types of service, but the backhand short service is used more often than other types of service in double~\cite{bwf:wilkieanalysis} and appears to get common in single since 2000s~\cite{waddell2000biomechanical}.
Consequently, most badminton players take considerable time practicing to perform an accurate and steady backhand short service.
Previous research works collected biomechanical data of professional players' or coaches' service motion~\cite{ahmed2011analysis, hussain2011videographical, phomsoupha2015science}.
These works focus on the stance at the moment of the contact without considering serving a shuttlecock is a continuous motion.


In this paper, we have implemented a self-training solution for practicing badminton backhand short service.
We analyzed the service motion with the characteristics of the open kinetic chain (OKC), which is a type of body movement in which the most distal segment is unconstrained and not fixed to anything~\cite{ellenbecker2001closed}.
These OKC characteristics thus are understandable parameters to visualize and help the user to adjust their continuous motion. 
We conducted a formative study to quantify an OKC model of backhand short service with six sub-elite players' motions and designed a training system that provides visualized feedback and guidance using the model.
Finally, we evaluated the system's effectiveness and usability by recruiting twelve users practicing with the system.
The result shows that our system can influence the user's exertion pattern toward our ideal model in short-term practice while a longer-term practice may be required to advance their service accuracy.

\begin{table*}[h!]
    \caption{
    The open kinetic chain characteristics of the backhand short service that we surveyed in this paper.
    We break down the motion of the backhand short service into these characteristics.
    }
    \label{tab:characteristics}
    \centering
    \begin{tabular}{llp{8cm}ll}
        \toprule
        \multicolumn{1}{c}{\textbf{Characteristics of motion}} &  & \multicolumn{1}{c}{\textbf{Backhand short service}}             &                                           & \multicolumn{1}{c}{\textbf{Open kinetic chain}} \\
        \cmidrule{1-1} \cmidrule{3-3} \cmidrule{5-5} 
        \textbf{Stress pattern}                                &  & Rotation of the wrist (adduction) and the elbow (supination)    & \multicolumn{1}{c}{$\longleftrightarrow$} & Rotary \\
        \textbf{Number of joint axes}                          &  & The craniocaudal (longitudinal) axis                            & \multicolumn{1}{c}{$\longleftrightarrow$} & One primary \\
        \textbf{Planes of movement}                            &  & Transverse plane                                                & \multicolumn{1}{c}{$\longleftrightarrow$} & Single \\
        \textbf{Number of moving joints}                       &  & The forearm remains stationary while the hand is moving         & \multicolumn{1}{c}{$\longleftrightarrow$} & One at a time \\
        \bottomrule
    \end{tabular}
\end{table*}

\section{Related Work}
\label{sec:relatedwork}
In this section, we review prior works related to our system: (1) the research in badminton, including coaching systems, motion analysis, and biomechanics, (2) motor skill acquisition and coaching in general, and (3) information visualization on feedback and guidance. 

\subsection{Research on Badminton}
Many researchers are dedicated to understanding Badminton in various fields, such as kinesiology, sport medicine, coaching, and sport engineering~\cite{bwfglobalresearch, phomsoupha2015science}.
In coaching badminton, Raina et al.~\cite{raina2017combat} presented a badminton training system that visualizes the user's muscular effort measured from wearable sensors and provides simple haptic feedback for the user to receive a real-time result of exertion pattern.
Sung et al.~\cite{sung2017implementation} implemented a badminton footwork training system using motion sensors and LEDs to analyze the player's motion and indicate the target position.
Kuo et al.~\cite{kuo2022improving} utilized a visual reaction training system to improve the player's agility and footwork.
He et al.~\cite{he2014development} designed an auxiliary system of motion synthesis with Kinect to help users learn badminton motions.

To help badminton athletes to gain deeper insights into their performance, more data and analyses have been collected and conducted for suggesting systematic training programs. 
Vial et al.~\cite{vial2019using} and Wilkie et al.~\cite{bwf:wilkieanalysis} defined an accurate service according to the anterior-posterior trajectory apex location and the vertical height of the shuttlecock when it crosses the net.
Rasmussen et al.~\cite{rasmussen2021simulation} simulate service trajectory with random speed, angle, and horizontal location to observe the influence of service height.
To assess the simulation result, they use area-of-attack, the area surrounded by the trajectory and the horizontal line through the top of the net within the opposing court, to classify those services.
In addition to the service, Waddell et al.~\cite{waddell2000biomechanical} investigated biomechanical research on badminton and summarized principles of performing a power shot, e.g. clear or smash.

In addition to the movement characteristics of the shuttlecock, there are works in kinesiology that focus on biomechanics or anthropometry~\cite{phomsoupha2015science}.
Ahmed et al.~\cite{ahmed2011analysis} and Hussain et al.~\cite{hussain2011videographical} collected and analyzed biomechanical variables like the joint angle of the arm and kinematic variables such as the speed of the shuttlecock through the video when recruited sub-elite players or coaches were executing backhand short service, forehand short service, and forehand long service.
Rusydi et al.~\cite{rusydi2015study} use three degrees of freedom inertia measurement units (IMU) placed on joints to measure the angle change and concluded that backhand short service could be classified as \textit{only use the wrist} and \textit{use both the wrist and the elbow}.

\subsection{Motor Skill Acquisition and Teaching}
Researchers have proposed training systems in various sports.
Wu et al.~\cite{wu2021spinpong} demonstrated how to design visual cues in virtual reality and introduced a pin-pong training system helping the user to face a rotating ball.
Han et al.~\cite{han2017my} applied virtual coaches surrounding the user in the augmented reality head-mounted display for learning Tai-Chi Chuan.
De Kok et al.~\cite{de2015multimodal} presented a coaching system that can automatically generate an instruction to a squat trainee.
Chan et al.~\cite{chan2010virtual} demonstrated a dance training system based on motion capture and virtual reality providing an interactive learning approach while practicing dance.
Zou et al.~\cite{zou2019evaluation} proposed a virtual-reality-based baseball batting practice system with real-time swing information.
Lin et al.~\cite{lin2021towards} introduced a basketball free-throw learning system and compared feedback in augmented reality display or 2D display.
Miles et al.~\cite{miles2012review} summarized the use of a virtual environment in ball sports training and pointed out that a virtual environment can provide consistent content, extra information, and instant change.

Besides, training motor skills other than sports are also highly attended.
Gould and Roberts~\cite{gould1981modeling} reviewed literature and identified influential factors related to motor skill acquisition.
Wulf et al.~\cite{wulf2010motor} summarized factors in motor skill training.
In addition to the theories, Ipsita et al.~\cite{ipsita2022towards} presented a virtual-reality-based welding training system attempting to solve the shortage of the welding workforce.
Maekawa et al.~\cite{maekawa2019naviarm} employed a wearable backpack-type haptic device with dual robotic arms to assist motor skill learning.

\subsection{Feedback and Guidance}
Some research works summarized a guideline for feedback design and visual design.
Blomqvist et al.~\cite{blomqvist2001comparison} compared two types of instruction in badminton training, and they found the user improved their service skill the most under the traditional instruction (perception, decision-making, and then movement execution).
Zhu et al.~\cite{zhu2020effects} proposed a guideline for using two types of feedback, knowledge of result and knowledge of performance.
Covaci et al.~\cite{covaci2015visual} analyzed the difference between first-person and third-person viewpoints while training free-throw with virtual reality.
Tang et al.~\cite{tang2015physio} and Semeraro et al.~\cite{semeraro2022visualizing} categorized visual designs for various types of movements.
Thoravi et al.~\cite{thoravi2019loki} reviewed the design space of remote teaching of physical tasks.
Most of our feedback and guidance design was inspired by these works.


\begin{figure}[!htb]
    \centering
    \includegraphics[width=\linewidth]{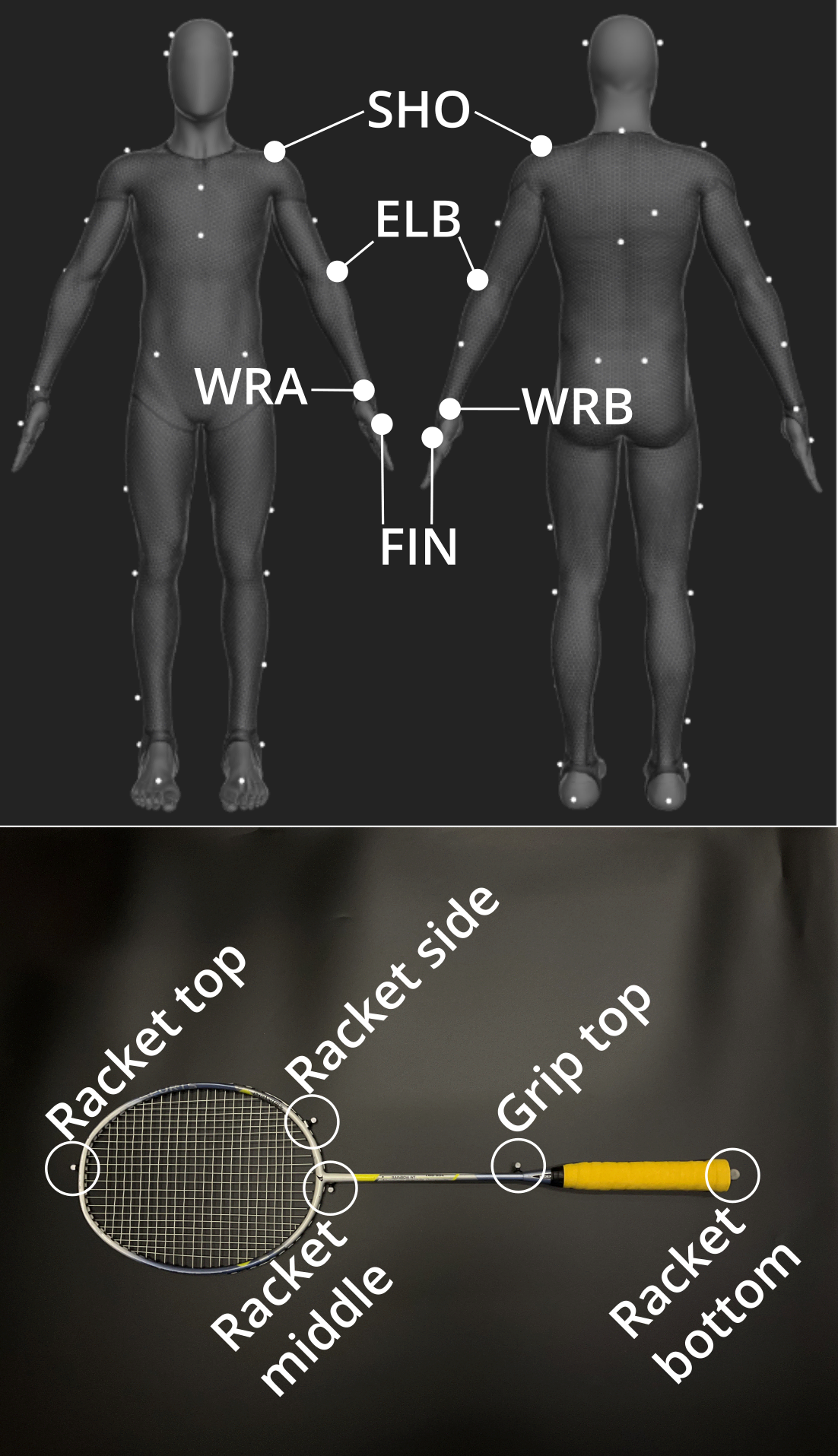}
    \caption{
    The conventional full body biomechanics markerset and the marker on the racket.
    The markers of joints on the arms are labeled as \textit{SHO}, \textit{ELB}, \textit{WRA}, \textit{WRB}, and \textit{FIN} with a prefix (\textit{L} or \textit{R}) indicating the direction.
    The actual position of the wrist is regarded at the middle of the \textit{WRA} and \textit{WRB}.
    According to the player's dominated hand, those markers are relabeled as \textit{hand}, \textit{wrist}, \textit{elbow}, \textit{shoulder}, \textit{shuttlecock-shoulder}, and \textit{shuttlecock-hand}.
    Finally, we derived target variables from the position of these markers (Appendix~\ref{app:kvdetail}).
    }
    \label{fig:marker}
\end{figure}

\begin{figure*}[!htb]
    \centering
    \includegraphics[width=\textwidth]{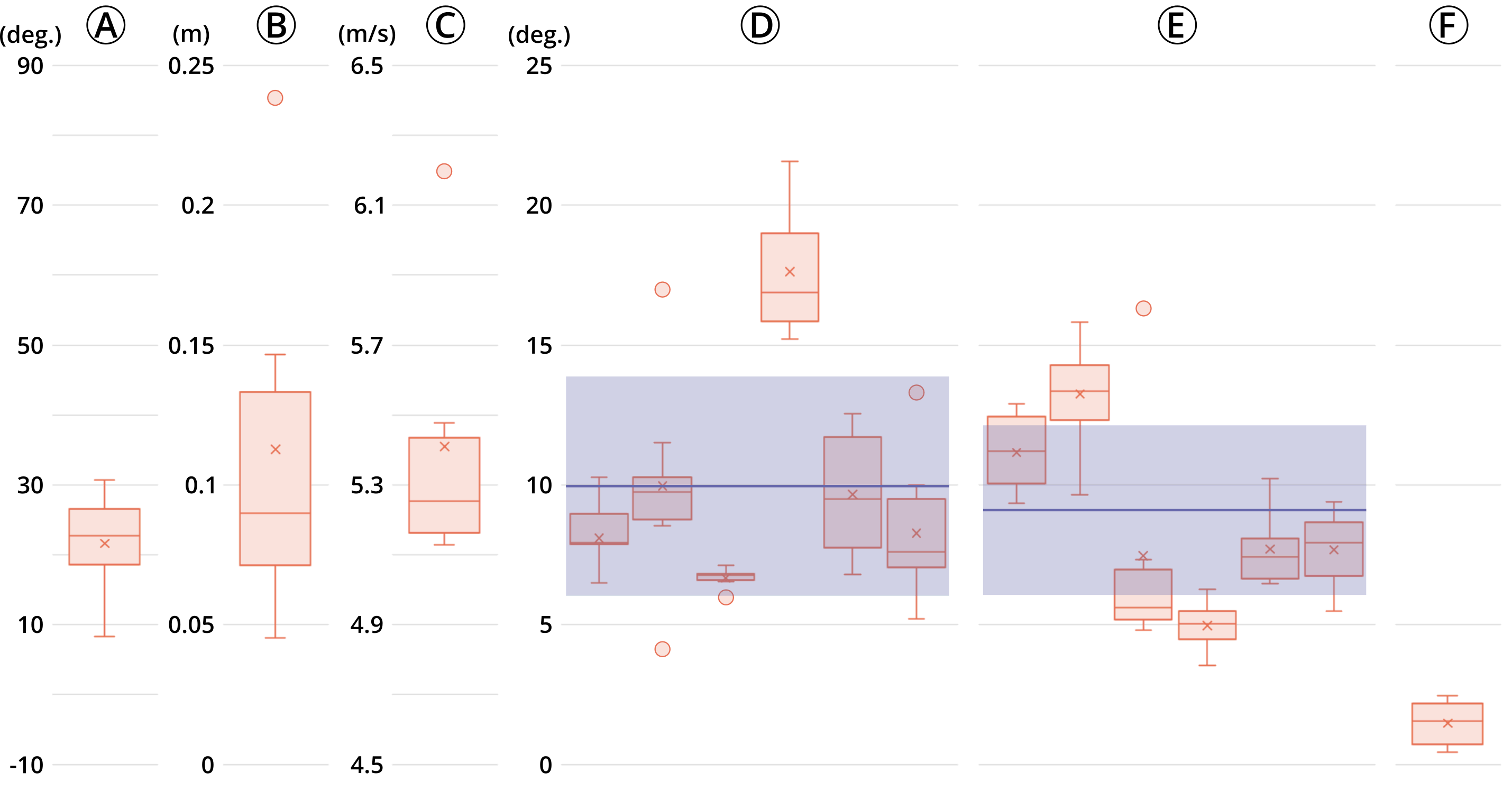}
    \caption{
    The distribution of six kinetic variables of interest from the formative study.
    (A) is the racket pitch angle, (B) is the change of the racket height, (C) is the racket speed, and (D) - (F) is the angle change of the wrist, elbow, and shoulder.
    The blue box in (D) shows the mean and standard deviation of all six players, whereas the blue box in (E) shows the mean and standard deviation of five players who use both the elbow and wrist.
    The cross mark denotes the mean, and the dot is the outlier.
    We found that P4 only uses the wrist during the study.
    We regarded P3 as the group using elbow and wrist because his elbow angle change is close to his wrist angle change, even though his elbow change is as little as P4.
    }
    \label{fig:formative}
\end{figure*}

\section{A Good Backhand Short Service}\label{sec:goodservice}

Understanding a good service is essential for designing a service coaching system.
An intuitive metric is whether the landing point locates in a valid region or not.
However, a legal service could be merely technically good, not tactically good, and results in the opponent's advantage.
Hence, previous works on the badminton service had come up with other metrics, such as the apex location and the clearance height of the trajectory, to quantify the quality of the service~\cite{vial2019using, bwf:wilkieanalysis}.

Another aspect of evaluating the service is the motion of the racket and the player's body.
In Section~\ref{sec:relatedwork}, we summarized works about the posture and the joint movement of the experienced player.
They provide kinetic variables, such as joint angles, racket angles, and the speed of the shuttlecock, helping the novice understand and reconstruct the posture.
Nevertheless, service is beyond the posture but includes four continuous steps: ready, back swing, forward swing, and contact~\cite{shen2014kinematics}.
Moreover, typically there is a follow-through movement after the contact.
Using the information of this \textbf{continuity between joints} on badminton service has not been fully explored. 

Inspired by the concept of the sequence of moving joints~\cite{waddell2000biomechanical}, we found the OKC, one type of body movement in which the furthest part away from the body is unrestricted and not fixed to any object~\cite{ellenbecker2001closed}, that is suitable to describe the motion of badminton service.
The motion characteristics of the OKC include stress pattern, the number of joint axes, muscular involvement, etc.
After investigating the OKC, we derived the mapping of backhand short service and OKC characteristics related to the motion process in Table~\ref{tab:characteristics}.

\subsection{Formative Study}
We conducted a formative study to quantify the OKC characteristics of the service and validate the mapping.
Besides, BWF adopted a new service rule that the whole shuttle shall be below 1.15m at the moment of contact in 2018~\cite{lawsofbadminton_2022}.
With the bygone service rule, the launch height depends on the player's rib height, which results in a taller player can serve flatter, i.e., it is harder for the receiver to attack.
Since many previous research works on service posture are earlier than the adoption, we also collected new data on kinetic variables through this study.

\subsubsection{Participants}
We recruited 6 sub-elite players from the badminton team of departments and the school, 1 female, 5 males, 1 left-handed, aged from 21 to 25 (M = 23.67, SD = 1.5).
Their experience of playing badminton range from 8 to 18 years (M = 10.5, SD = 3.83), and the duration under systematical training varies from 1 to 9 years (M = 4.5, SD = 3.08).
In addition, P5 is the gold medalist of 3 city individual tournaments in double, P1 is the champion of an intercollegiate team's competition, P2 is third place in an interdepartmental contest, and P4 is the runner-up in double and third place in single in interdepartmental competitions.

\subsubsection{Apparatus, Procedure, and Task}\label{sec:formativeapparatus}
The study was conducted at Human Computer Interaction Lab at National Taiwan University with markings of part of the short service line and center line on the floor. 
We used six OptiTrack Prime 13 cameras to capture the motion of players.
Markers were applied based on the OptiTrack pre-defined conventional full-body biomechanical markerset (Figure~\ref{fig:marker}).
Also, we placed markers on the racket to monitor its speed and orientation.
The shuttlecock, however, was not tracked because the marker or the stained spot on it was not always visible to the camera of a retro-reflective motion capture system due to its slant shape~\cite{rasmussen2021simulation}.

We recorded one player's motion at a time.
In the beginning, they were informed of the task details and asked to put on the motion capture suit.
Participants had 5 minutes for warm-up to get familiar with provided racket while wearing the motion capture suit.
After the warm-up, they were asked to perform 20 to 40 services, ensuring at least five successful and accurate services without losing marker information were recorded.
Each recording starts from the moment of the backswing and terminates as soon as the racket contacts the shuttle.
Afterward, they took off the suit and filled out the form of their badminton experience and personal information. 

\begin{table}[!htb]
    \caption{
    Ideal kinetic variables of backhand short service.
    These variables include the motion of the racket and the body.
    Both backhand short service patterns were used in the formative study and those two types of shots were recorded in different elbow angle changes.
    }
    \label{tab:kineticVariableResult}
    \centering
    \begin{tabular}{rl c c l}
        \toprule
        \multicolumn{2}{c}{\textbf{Kinetic variable}}                   & \textbf{Mean}         & \textbf{SD}           & \multicolumn{1}{c}{\textbf{Remark}}   \\
        \midrule
        Racket pitch angle                  & (deg.)                    & 21.60                 & 7.95  \\
        Racket height difference            & (m)                       & 0.11                  & 0.07  \\
        Racket speed                        & (m/s)                     & 5.41                  & 0.41  \\
        Wrist angle change                  & (deg.)                    & 9.96                  & 3.93  \\
        \multirow{2}{*}{Elbow angle change} & \multirow{2}{*}{(deg.)}   & \color{violet} 9.10   & \color{violet} 3.04   & \color{violet}\small Elbow and wrist   \\
                                            &                           & \color{blue} 4.97     & \color{blue} 0.96     & \color{blue}\small Wrist only    \\
        Shoulder angle change               & (deg.)                    & 1.48                  & 0.87  \\
        \bottomrule
    \end{tabular}
\end{table}

\subsubsection{Data Analysis and Result}
The captured motion was stored with a two-dimensional array whose axes are timestamps and each marker's three-dimensional coordinate.
With the distance between the racket and the hand holding the shuttle, we labeled keyframes, including the start of the backward swing, the start of the forward swing, and the moment of contact.
Next, we derived segments and kinetic variables, such as joint angles, racket angle, and racket speed, from the change of the marker position in each period (Appendix~\ref{app:kvdetail}).
In addition to these variables, we also observed the racket height during forwarding swinging and joint angle change during the whole swinging process by calculating the difference.
Since the performance of a service depends on the trajectory of the shuttlecock, and we lacked the information about the shuttlecock, we put the racket motion to use instead.
At the same time, considering OKC characteristics, we summarized variables affecting the shuttlecock and related to the body motion, including the racket pitch angle, the racket height change, the racket speed, and the involvement of the wrist, elbow, and shoulder in Figure~\ref{fig:formative}.
After removing outliers by applying the 1.5 $\times$ interquartile range rule, the mean and standard deviation of these variables of a good badminton backhand short service is shown in Table~\ref{tab:kineticVariableResult}.

\begin{figure}[!htb]
    \centering\includegraphics[width=\linewidth]{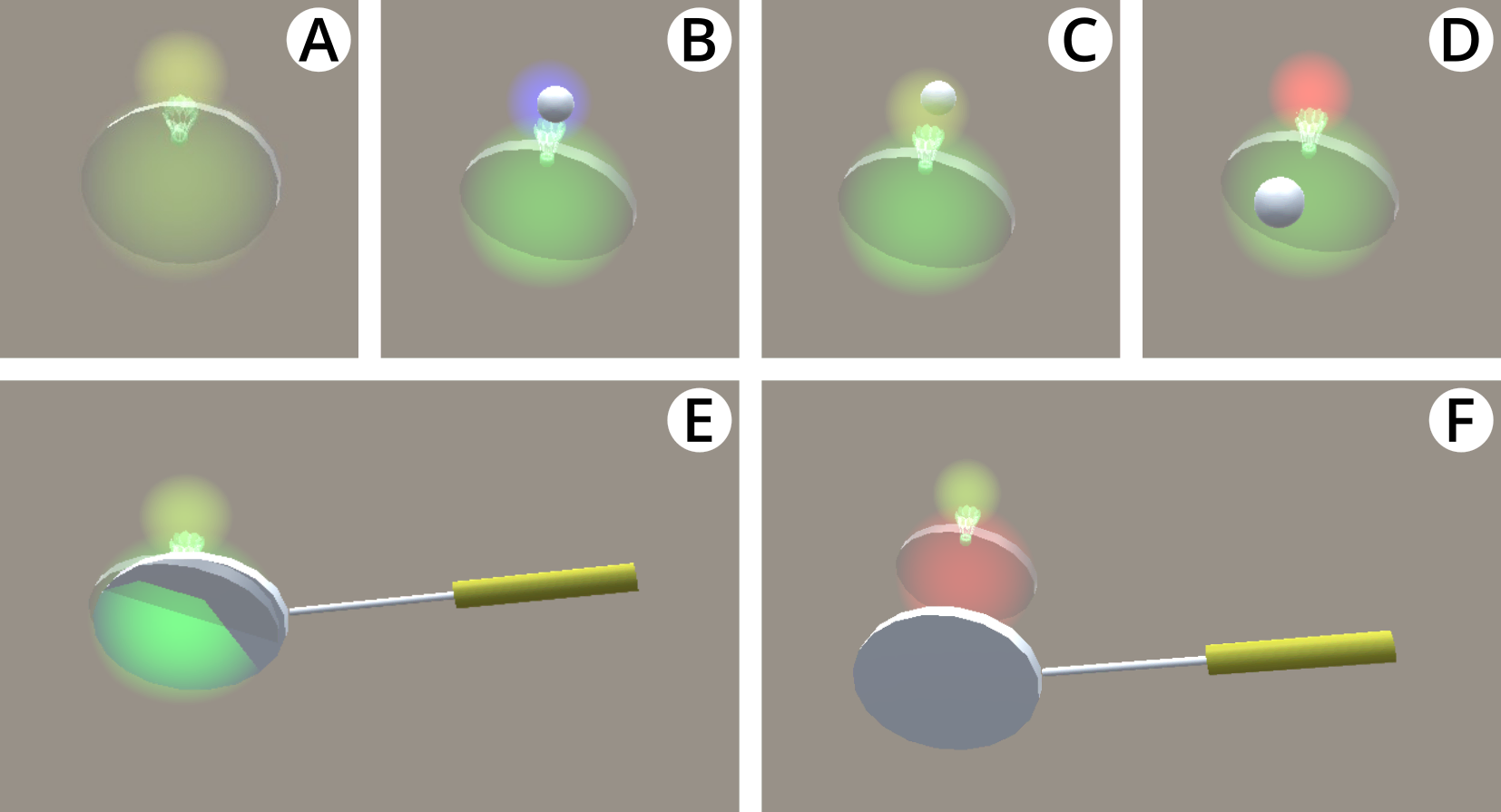}
    \caption{
    The guidance of the ready position of the shuttlecock and the racket.
    The position is real-time generated according to the user's stature.
    (B) - (F) shows the color change depends on the distance between the actual position and the target position of the shuttlecock and the racket along the sagittal axis of the user's body.
    }
    \label{fig:ready}
\end{figure}

\begin{figure}[!htb]
    \centering
    \includegraphics[width=\linewidth]{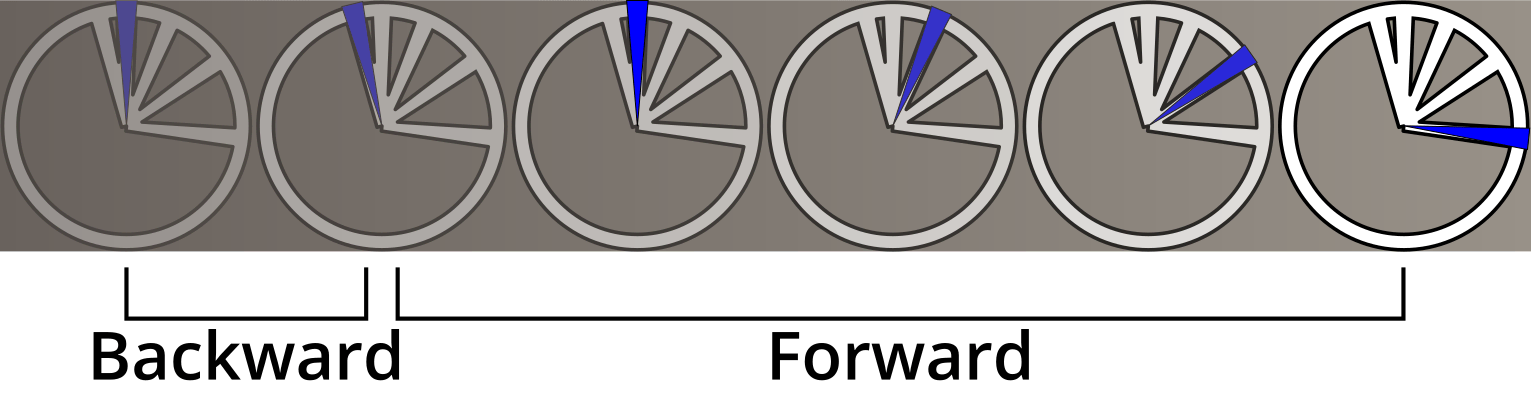}
    \caption{
    The frames of the animation of the swing track guidance.
    The indicator swipes backward first and then forward.
    }
    \label{fig:swingtrack}
\end{figure}

\begin{figure}[!htb]
    \centering
    \includegraphics[width=\linewidth]{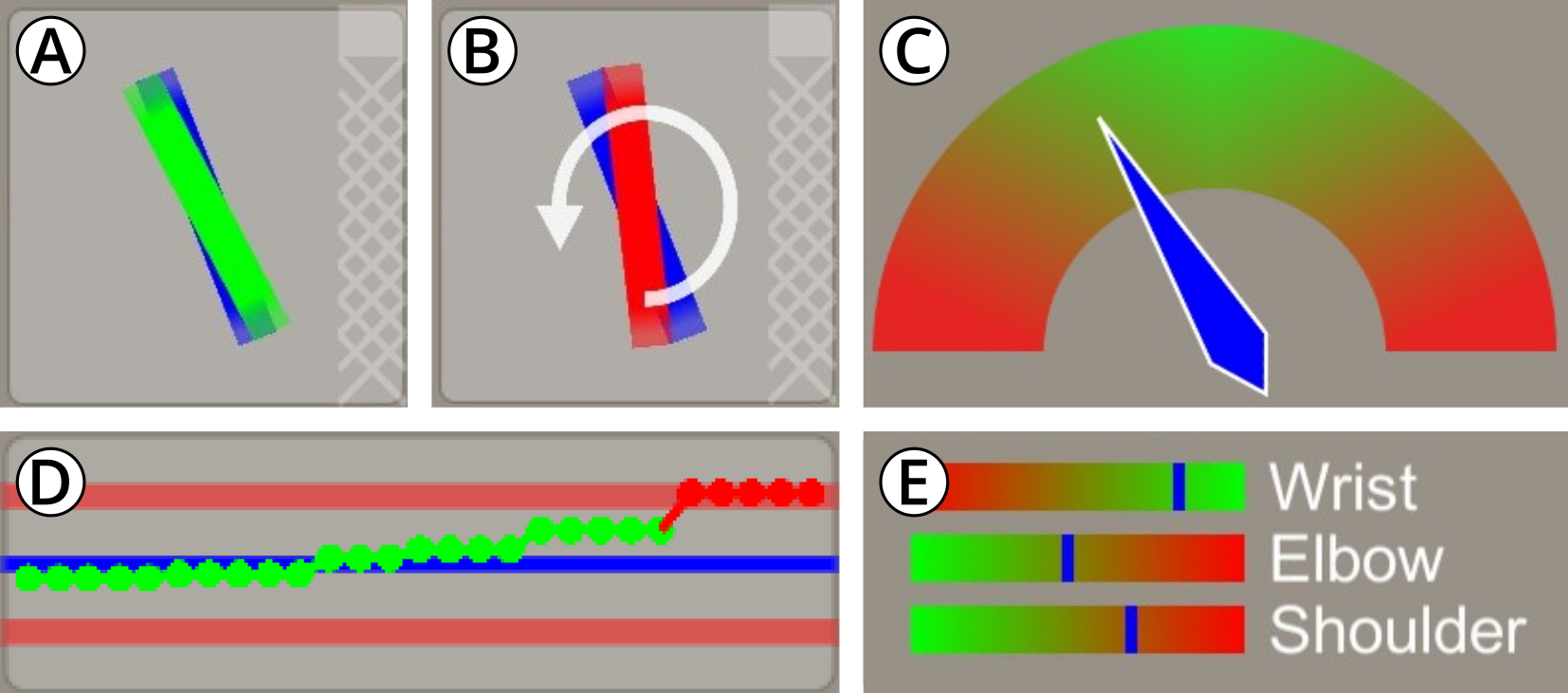}
    \caption{
    (A) and (B) illustrate the racket pitch angle difference between the shot performed by the user, and the recommended angle is within or without the threshold.
    The resulting angle exceeding the threshold is colored red with a round arrow indicator.
    (C) depicts the racket speed at the moment of contact with a speedometer whose middle represents the target speed.
    (D) depicts the height change of the racket relative to the ready position of the racket during the forward swing.
    If the height difference is greater than the standard, it is recorded in red; otherwise, in green.
    (E) shows the joint angle change relative to the target motion pattern.
    The wrist should be close to the right side (larger side), while the other two joints' usage must be kept on the left side (smaller side).
    }
    \label{fig:feedback}
\end{figure}

\section{System}
With the quantified result of kinetic variables of the service, we designed a self-training system utilizing Unity and providing visual guidance and real-time feedback.
The system focused on the motion of serving only with the wrist to lower the bar on learning the service and simplify its design.
This system comprises three parts: motion capture, motion analysis, and feedback.
It is possible to use any motion capture method, and we use the same approach as the formative study (Sec.~\ref{sec:formativeapparatus}) here.
Meanwhile, the captured data is streamed and analyzed in a Unity scene.
Finally, the analyzed result is visualized as the guidance and feedback shown on the screen in front of the user as Figure~\ref{fig:figure1}D shows.




\subsection{Motion Analysis}
While the captured skeletal data was real-time streamed to the Unity scene and bound to an avatar, there is a kinetic variable calculator that extracts variables listed in appendix~\ref{app:kvdetail} from the avatar.
Next, there are two types of components, the state machine of service state and the controller of guidance and feedback.

As addressed in Section~\ref{sec:goodservice}, we separate a service into five states: ready, backward swing, forward swing, contact (follow through), and idle.
A service state machine monitors the user's kinetic variables and updates the service state.
Starting from the idle state, if the user holds the racket and the shuttle close enough to the calculated ready position shown on the screen, the state transfers to the ready state.
In the ready state, the state machine transfers the state into the backward swing state as soon as the racket starts moving in the opposite direction of the body's forward.
Otherwise, the state will switch back to the idle state immediately if the user moves the racket in another direction (e.g., forward or upward) while ready because the backward swing could help the user to exert easier to maintain better consistency between short and long service in the future.
When the distance between the racket and the shuttlecock gets closer in the backward swing state, the state changes into the forward swing state.
Right after the distance increases, the state turns into the contact state and switches back to the idle state a few seconds later.

As the state changes, there is ready position and swing track guidance and feedback on the service result, the variable difference between the user's attempt and the expert model.
At the beginning and the end of every shot, there are translucent indicators of the shuttle and the racket with halos to illustrate a proper ready position.
After the user holds the shuttlecock and the racket at the ready position, the position indicators disappear, and a circular indicator with animation shows up to indicate the swing track.
As soon as the user swings, all the guidance and feedback are hidden until the user hits the shuttlecock.
Afterward, the feedback revealing the difference between the current trial and the expert model is shown on screen.

\subsection{Guidance and Feedbacks}
The guidance informs the user where (the ready guidance) and how (the swing guidance) to perform a backhand short service.
In consideration of the BWF adoption, the height of the target shuttlecock position is set at a constant height, and the position in the transverse plane depends on the user's front and shoulder position.
As Figure~\ref{fig:ready} shows, the color of the halo indicates the distance between the suggested position and the actual position in the sagittal axis (body's front).
Once in the ready state, a circle with spokes is displayed to indicate the swing track (Figure~\ref{fig:figure1}C).
The center of the circular indicator is the user's wrist, and the plane of the circle aligns with the transverse plane through the shuttlecock.
The angle between adjacent spokes increases with a constant angular acceleration calculated from the user's ready posture and the target racket speed.
To make it easier to understand, we animated the change in spoke color and scale to indicate the motion of a backward and then forward swing (Figure~\ref{fig:swingtrack}).

The feedback contains the motion of the racket and the body.
The motion of the racket directly affects the flight of the shuttlecock and the motion of the body is exactly what we are going to teach the user.
We used the racket pitch angle, the height change of the racket, and the racket speed to describe the movement of the racket.
In the meantime, we illustrate the user's joint involvement to indicate his/her exertion pattern.

After the shot, all four types of feedback are shown on the display as Figure~\ref{fig:figure1}A illustrates.
The racket pitch angle feedback indicates the angle at the moment of contact.
The target angle is shown as a tiled blue rectangle, and the result of the user is displayed with either green (Figure~\ref{fig:feedback}A) or red rectangle (Figure~\ref{fig:feedback}B) depending on the difference between the target and what the user performs.
If the difference is greater than the standard deviation of the expert model, the color is red and a circular arrow towards the desired angle pops up.
The height change of the racket from the start of the forward swing to the moment of contact is presented with a line chart (Figure~\ref{fig:feedback}D).
The baseline at the center represents the height at the start of the backswing, and the user needs to keep the change within one standard deviation of the ideal model.
The height difference within the threshold is colored green. 
Alternatively, red while it exceeds the bound.
The racket speed at the moment of contact is displayed as an analog speedometer (Figure~\ref{fig:feedback}C).
The user is asked to keep the pointer in the middle.
The joint involvement is graphed as three sliders (Figure~\ref{fig:feedback}E).
The difference between maximum and minimum angles during the swing is illustrated as the blue indicator in the slider.
The middle of each slider represents the average of each joint usage of the model.
The target motion of our design is a service using the wrist only.
Thus, we expect that the change in the wrist angle should be as significant as possible.
On the contrary, the other two joints should be fixed during the swing.
Since the angle between the shuttlecock and the racket at the end of the backward swing can be less than the model's average, we use the small one as the upper bound of the wrist involvement judgment.

\begin{figure}[!htb]
    \centering
    \includegraphics[width=0.9\linewidth]{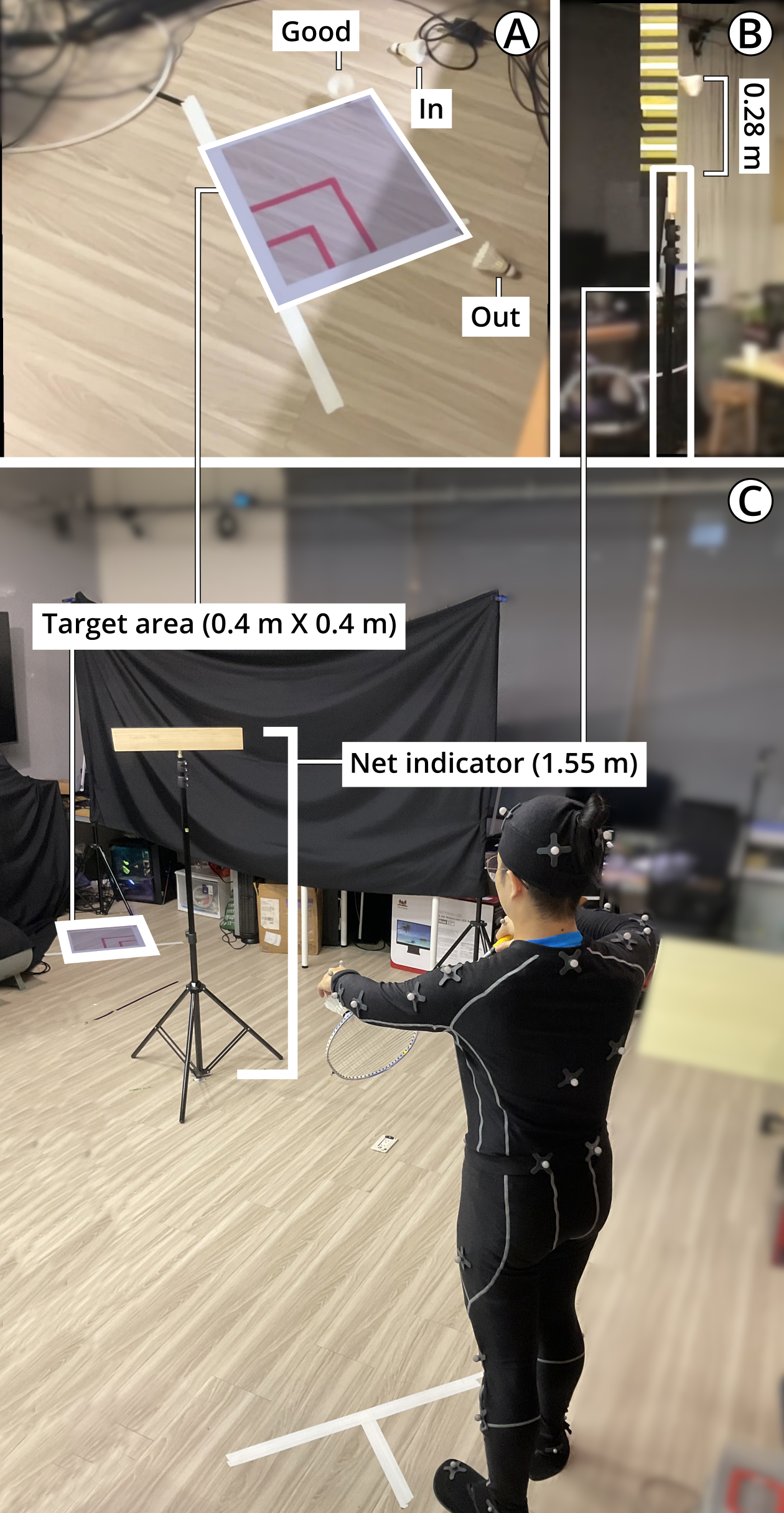}
    \caption{
    The overview of a test session.
    (A) shows a shuttlecock that just lands and the target region, a 40 cm $\times$ 40 cm square next to the service line and the center line.
    A shuttlecock lands in the target area is labeled as \textit{Good}, the other valid part as \textit{In}, and the rest as \textit{Out}.
    (B) is the side view of the net indicator while a shuttlecock passes through.
    We check the apex location and if the shuttlecock successfully passes through the net indicator.
    There is a board with stripes spaced 2 cm wide above the indicator for measuring the clearance height after the test.
    (C) presents a user that is serving in the test session.
    There are markings of the short service line and center line and an indicator of the net.
    The width and distance of the marking are the same as the laws of badminton promulgated by BWF.
    The user is requested to serve to the opposing right service court marked on the floor.
    }
    \label{fig:test}
\end{figure}
\begin{figure}[!hbt]
    \centering
    \includegraphics[width=\linewidth]{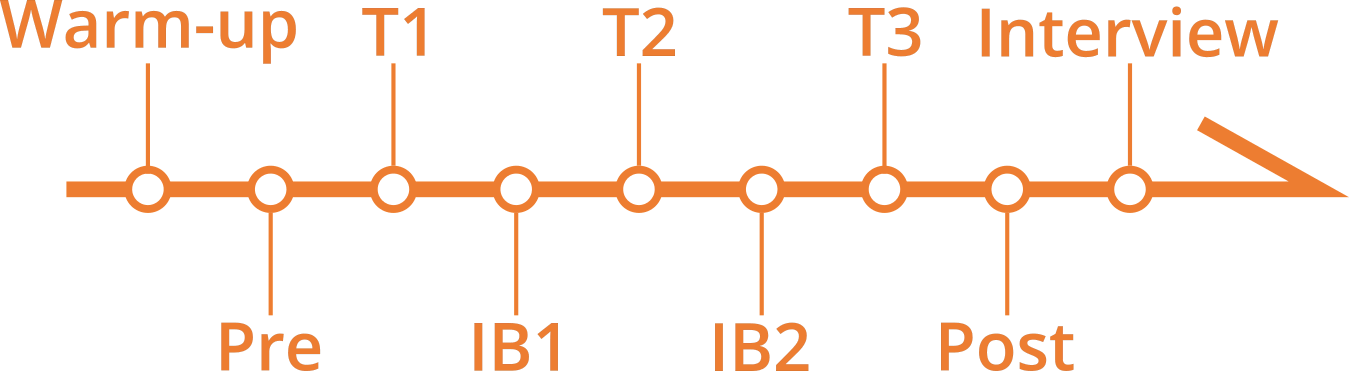}
    \caption{
    The timeline of the evaluation study.
    \textit{T1} $\sim$ \textit{T3} denote the first to the third training session.
    \textit{Pre} denotes pre-training test, \textit{IB} denotes in-between test, and \textit{Post} denotes post-training test of evaluation study.
    }
    \label{fig:timeline}
\end{figure}
\begin{figure}[!hbt]
    \centering
    \includegraphics[width=\linewidth]{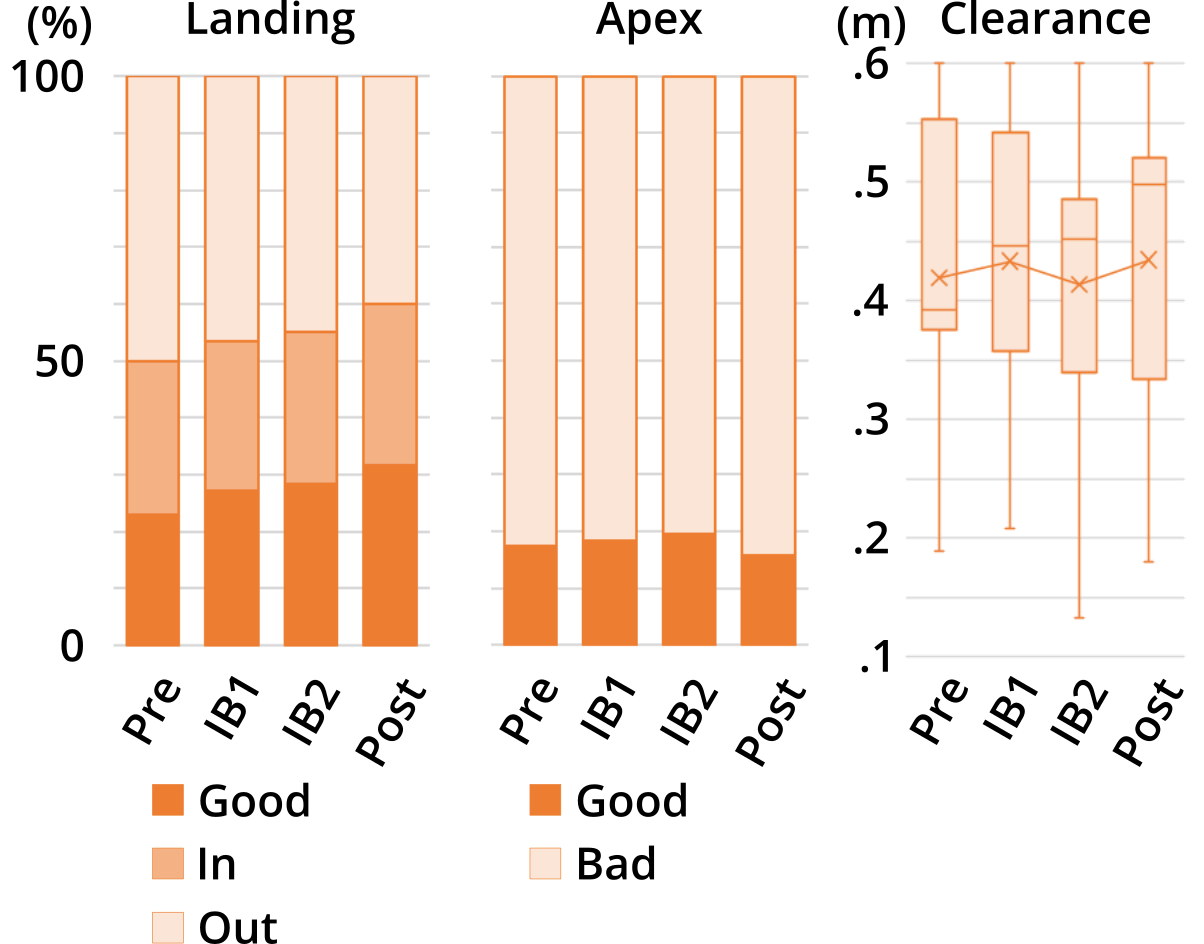}
    \caption{
    The measurements of the shuttlecock trajectory in four test sessions.
    \textit{Pre} is pre-training test, \textit{IB} is in-between test, and \textit{Post} is post-training test of evaluation study.
    The left and the middle chart shows the percentage of each category (landing type and apex type) of shots performed by a user on average, and the right chart illustrates the distribution of the average clearance height of every user.
    The cross mark denotes the mean.
    }
    \label{fig:trajectoryresult}
\end{figure}

\begin{figure*}[!htb]
    \centering
    \includegraphics[width=\textwidth]{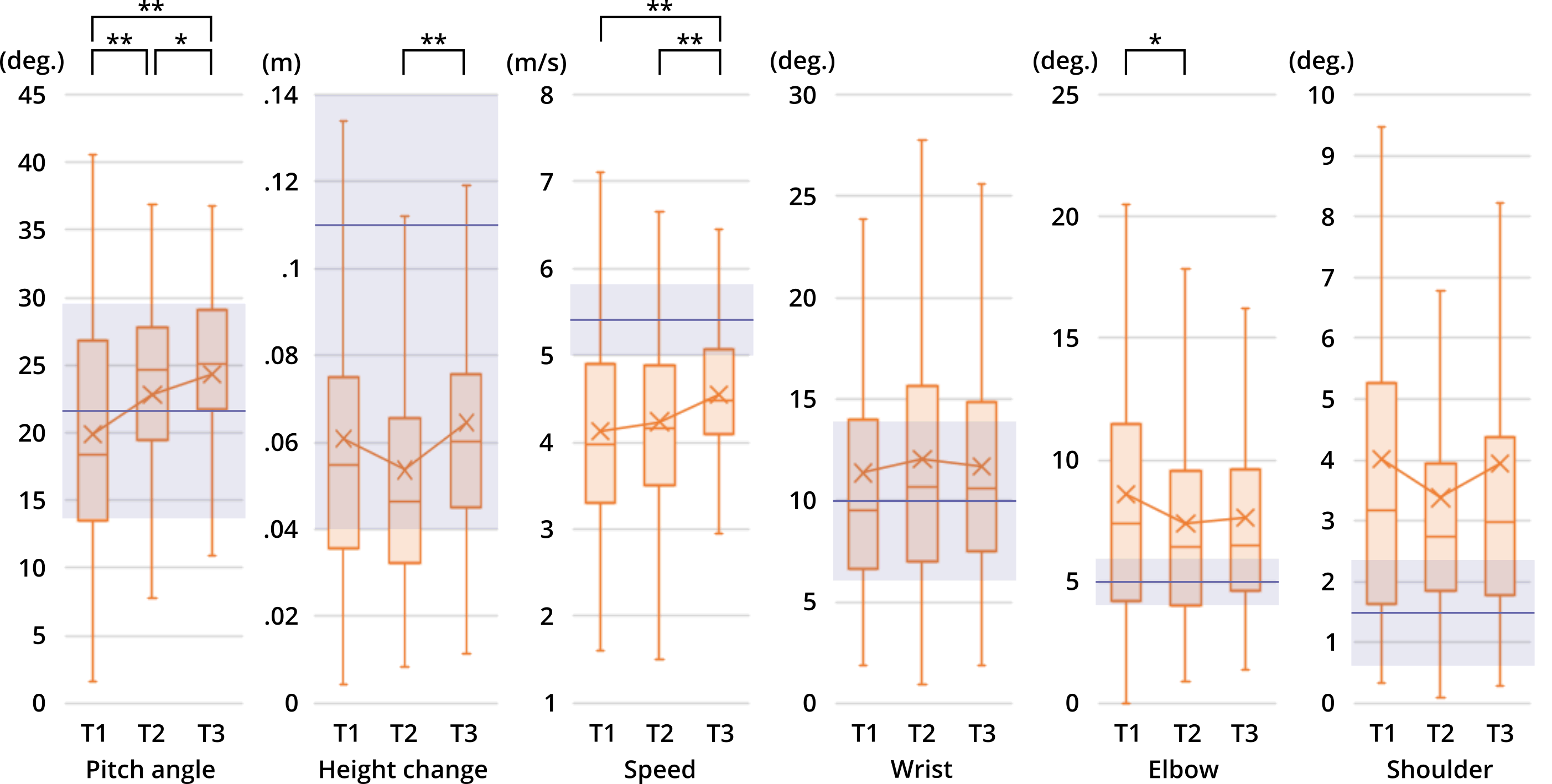}
    \caption{
    The service motion of all the users during three training sessions.
    T1, T2, and T3 denote the first, the second, and the third training session.
    The cross mark denotes the mean.
    The dark line and the colored region are respectively the averages and the standard deviation of each variable from the formative study (Table~\ref{tab:kineticVariableResult}).
    In most trials, the user can keep the racket pitch angle and the height change within one standard deviation from the average.
    Also, the user can better master the racket speed after training.
    Since we promote using only the wrist in the system, wrist involvement remains above the norm.
    The involvement of the elbow has a trend toward the line, while the shoulder involvement has no significant difference.
    But note that the mean and the median of shoulder involvement are less than five degrees which is similar to the standard of the elbow, the other fixed joint.
    Except for the wrist and shoulder involvement, other variables have statistical significance between sessions by pairwise t-test and more concentrated distribution.
    *: \textit{p < 0.05}, **: \textit{p < 0.01}
    }
    \label{fig:userkvresult}
\end{figure*}

\begin{figure}[!htb]
    \centering
    \includegraphics[width=\linewidth]{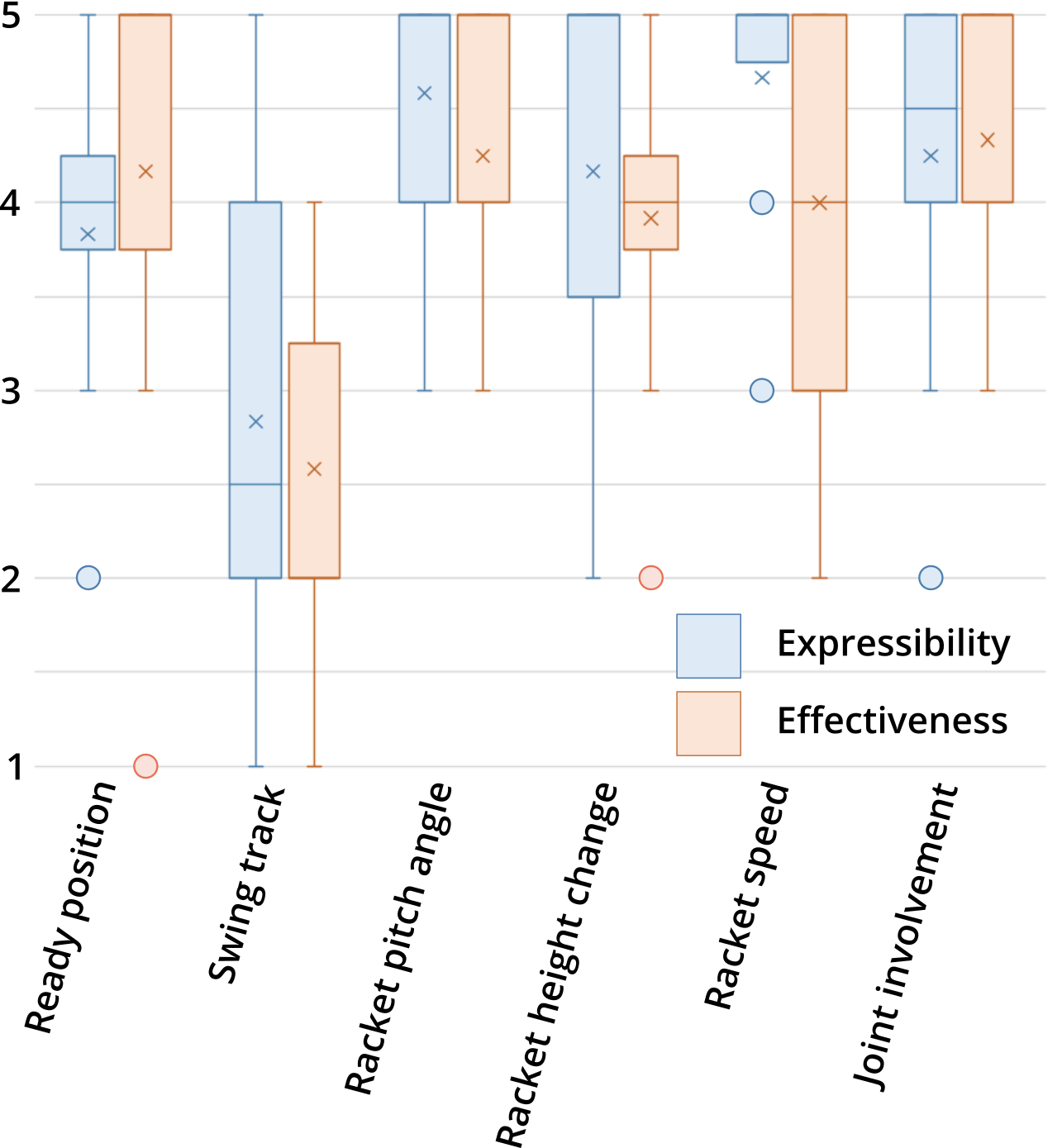}
    \caption{
    The response to the visual design questionnaire: five-point Likert scales of visual design's expressibility and effectiveness during training.
    The cross mark denotes the mean, and the dot is the outlier.
    Most of the designs received positive feedback except for the swing track guidance since this design could not reflect the user's motion in real-time, while the user can compare his/her motion with the ideal model by checking the immediate change of other designs.
    }
    \label{fig:interview}
\end{figure}

\section{Evaluation}
The evaluation of our system includes the usability and the improvement of the user's performance.
We conducted a study on practicing backhand short service with our system.
We measured the performance by analyzing the change in the user's kinetic variables and the trajectory of the shuttlecock before and after training.
Lastly, we interviewed the user to evaluate the system's usability and collect their subjective feedback about the visual guidance.

\subsection{Participants}
We recruited 12 users, 3 females, and 9 males, all right-handed, aged from 19 to 27 (M = 22.83, SD = 1.95).
Their total experience of playing badminton is less than five years (M = 2.14, SD = 1.53).
Most of their training experience comes from the physical education class while they were schooling, and the duration is less than three years (M = 1.12, SD = 1.11).

\subsection{Apparatus, Procedure, and Task}
The study was also conducted at Human Computer Interaction Lab at National Taiwan University. 
We utilized a TV (Samsung UA65RU7400WXZW) to display the visual guidance and feedback, a PC (i7-10700KF, 32G RAM) to analyze the user's motion, and the same OptiTrack setup as the formative study to capture the user's motion.

Figure~\ref{fig:timeline} shows the timeline of our study procedure.
The user was brought to the lab and practiced one by one.
The study consisted of a pre-training test session, three training sessions, two in-between test sessions among training, a post-test session, and an interview on system usability and subjective feedback on visual design.
The user was requested to follow the service line marking and serve twenty shots from the right service court to a 40 cm $\times$ 40 cm square next to the center line and the short service line in each test session (Figure~\ref{fig:test}C).
The trajectory was recorded by two mobile devices (iPhone 7 and iPad mini 6) in 1080p 240 fps slow motion mode.
With Matlab, we calibrated the camera and extracted the keyframes, including the moment of passing through the net and landing.
In each training session, the user was asked to practice fifty shots in front of the screen and pay attention to the given feedback instead of the trajectory of the shuttlecock.
There was a short interval (thirty to sixty seconds) between every ten shots and a short break (one to three minutes) in the middle of two sessions.
At the start of the study, every user had about five minutes to warm up and then performed twenty shots for the pre-training test.
Following the pre-training test, the user was asked to put on the motion capture suit and start the training session.
All the training sessions and in-between test sessions were alternated.
The user moved on to the post-training test while finishing all the training.
After the last test session, the user took off the suit and had an interview with the experimenter.
During the interview, the user was demanded to report feedback on the system and fill in a questionnaire comprising the standard system usability scale (SUS)~\cite{sus} and a five-point Likert scale (1 = strongly disagree, 5 = strongly agree) about "Can the visual design convey the following content?" and "Does this guidance/feedback help you while training?"
The study took about ninety minutes, and the user was compensated with 450 NTD.

\subsection{Result}
The evaluation result contains the measurement of the user's performance and the outcome of the interview.

We chose the landing point, the clearance height, and the apex location to measure a shuttlecock trajectory.
We measured the metrics from processed frames and summarized them in Figure~\ref{fig:trajectoryresult}.
As Figure~\ref{fig:test}A illustrates, the performance of the landing point is classified into three categories: \textit{Good} (within the 40 cm $\times$ 40 cm square), \textit{In} (within the other valid region), and \textit{Out} (others).
From the viewpoint of Figure~\ref{fig:test}B, an apex locates between the net and the user is labeled as \textit{Good}; otherwise, \textit{Bad}.
The clearance height is measured by two-centimeter-wide stripes attached to the net indicator (Figure~\ref{fig:test}B).
Though most of the shuttlecocks pass through the indicator in the range of board width, there is still measurement error because of the distance between the shuttlecock and the ruler.
This error can be calculated by the similar triangles formed by the camera, the shuttlecock, and the stripe board.
In the current setup, the error could be up to $\frac{1}{3}$; hence, the primary function of the clearance height is to determine whether a shot succeeds in passing the net.
Finally, we found no significant improvement in service accuracy with a paired-sample t-test.

The user's motion was recorded with Unity and dumped as the result of the formative study.
We wrote Python scripts to parse the records and derive the kinetic variables at every moment (Appendix~\ref{app:kvvariation}).
We found jitters in some trials and labeled these shots according to the number of local extreme values.
After deducting trials that were jitters or lost tracking, the minimum common number of valid services is twelve among sessions.
We summarized all users' kinetic variables of each session's first twelve valid services in Figure~\ref{fig:userkvresult}.
Compared with a pairwise t-test, there is statistical significance in the racket pitch angle, height change, swing speed, and the involvement of the elbow.
Furthermore, the result of these four variables became concentrated progressively.
The mean and median of the racket pitch angle, height change, and wrist involvement keep within a standard deviation from the average of the expert model.
Their swing speed gradually increased toward the normal of the ideal model.

The average SUS score was 70.42 (Min = 50, SD = 14.41, between \textit{Good (71.4)} and \textit{OK (50.9)} in the adjective rating~\cite{susrating}).
And the score of the questionnaire about the visual design is shown in Figure~\ref{fig:interview}.


\section{Discussion}
The following discussion has two directions: the interpretation of the study result and the characteristic of badminton backhand short service.

\subsection{Study Result}
Though the users' performance in shuttlecock trajectory had no significant improvement in such short-term training, the number of shuttlecocks landed within the target square slightly increased on average as Figure~\ref{fig:trajectoryresult} shows.
At the same time, the users were able to maintain the racket pitch angle, height change, and wrist involvement close to the ideal model.
Moreover, many of them felt that their exertion patterns were affected.
As we promoted using only the wrist in our system, their elbow usage continuously decreased and got concentrated, while their wrist involvement remained above the standard in most trials.
Although the shoulder involvement had no significant difference, its means and median in each session were less than five degrees which was similar to the elbow involvement and relatively minor to the use of the wrist.
In the meantime, the more concentrated distribution of the racket kinetic variables and the elbow involvement indicates that the user's motion became more consistent.
These results can be observed from the change among training sessions depicted in Figure~\ref{fig:userkvresult}.

As Figure~\ref{fig:interview} illustrates, most visual designs were thought to be expressible and effective except for the swing track guidance.
Though some users rated the swing track guidance up to five on expressibility, the guidance did not catch most of the users' attention while training.
This outcome might result from the poor interactivity of the guidance, i.e., this guidance could not reflect the user's motion immediately.
In contrast with other designs, the user could have real-time feedback while attempting to align the ready guidance (Figure~\ref{fig:ready}) or after the shot (Figure~\ref{fig:feedback}).
Furthermore, the swing track could be decomposed as the racket height change and the swing speed.
Thus, the swing track guidance was only necessary to inform the user to start the swing.
A possible solution is to render the racket position for keyframes of swinging.
The user could distinguish the difference between the trial and the track with the new feedback after serving, though the user still can not rectify the motion while swinging.
In summary, the coaching system designer must pay attention to the interactability of feedback.

The next critical issue is the cost of the motion capture.
Most replies to SUS are positive except for the question about the need for a technical person.
Due to the current motion capture setup, the user needs to wear markers; in the meantime, some markers should be placed on the user's back.
Moreover, the skeletal model build or recalibration requires the user to maintain still.
Consequently, at least one person that knows the initial motion capture environment setup is needed.
In order to improve the rating, we may adopt an alternative motion capture solution to lower the cost of use.

There are few negative effectiveness reviews of ready guidance, racket height change, and racket speed feedback.
According to the interview, if the user's exertion habit could fit the system standard before the training, they tended to ignore the variables because too much information needed the users' attention within a trial.
Users also reported that it was hard to focus on all feedback once in a shot.
For instance, to serve faster, they unconsciously utilized the power from the upper arm by moving the elbow and shoulder.
An intuitive solution is displaying less feedback after serving or highlighting feedback based on its importance.
Both two approaches depend on the feedback priority can be determined by the match rate of each feedback or the user's decision.

Besides, the users' prior badminton knowledge was essential in their training experience.
Users with less knowledge could have questions about the instruction and have no point of contact to answer their confusion.
On the contrary, an experienced user may be used to serve differently, but the system cannot examine the correctness of all the service methods.
A user with another proper service approach does not need to employ this system.
A user with an improper service habit may take more time to rectify.
Their demand for the system are contrasting, but they may not be able to judge by themselves.
Therefore, a clear statement of the purpose and target audience of a self-training system for a user to appropriately utilize the system is critical.
And a more thorough study of motion models is necessary.

\subsection{Nature of Backhand Short Service}
Some users reported that the swing speed feedback can not precisely respond to their motion occasionally.
Except for the losing tracking of motion, there are other possible reasons.
One is the difference between the rotary and translational motion of the racket.
Given a constant force and acceleration time, the speed on the top of the racket in rotary motion is greater or equal to that in translational motion (Appendix~\ref{app:racketspeed}).
Another reason is the duration of the follow-through motion.
If the user stopped swinging right after the contact, the racket speed might decrease before the contact.
This resulted in the sampled speed might be lower than the user thought.

The lack of the user's lower body information bothered some users.
These users are much taller than others and the sub-elite players in the formative study.
They reported that they needed to do a quarter squat to get a more comfortable arm stance, but the system did not provide information on the lower body.
This made them doubt their posture until they could match all the feedback.
To solve the problem, we can integrate the posture model summarized by prior research works.
The height limit in our system and the proper arm position provided by previous works make the user bend the knee naturally.
This posture resembles the tall professional players' service stance in international tournaments.

Another issue is about joint involvement.
For instance, the shoulder movement has two main directions, flexion/extension and abduction/adduction.
The former influences the swing direction and the pitch angle more than the latter.
However, our system only reports the general change of the upper arm without factoring out the two components of angles.
Moreover, all three arm joints have more than one degree of freedom which results in the user's confusion about how to rectify the joint usage.
Consequently, it is vital to clarify all the detail of each joint with an appropriate priority.


\section{Limitation and Future Work}
In this section, we document our system's limitations during the evaluation study and sketch our expectations for the system.

The trajectory of the shuttlecock depends on not only the contact speed but also the contact angle.
Without tracking the shuttlecock, we could not provide instructions on the shuttlecock's orientation and the contact point on the racket.
As some users reported in the interview, sometimes the flight of the shuttlecock was obviously unacceptable, while all the metrics shown on the screen were matched.

The other issues were related to the motion capture.
In the current setup, the user's skeletal model might drift occasionally.
The drifting resulted in the wrong analysis fallout and the inconvenience of recalibrating.
Moreover, some users mentioned the discomfort of wearing the motion capture suit in the interview.

Besides utilizing more motion capture cameras to increase the system stability, it is also an option to apply other tracking technologies.
Wearable IMU-based sensors have been advocated as a cost-effective solution for tracking motion in unsupervised situations like home~\cite{komaris2022unsupervised}.
We can also track the shuttlecock with a deep learning network~\cite{huang2019tracknet}.
Eventually, the system can be deployed on a commercial device, such as Kinect or Rokoko, to make it more inexpensive and comfortable.
In addition to motion capture, exhibiting the motion model with a head-mounted display in virtual or augmented reality could make it easier to comprehend.
At the same time, the motion model could be more than a single-joint comparison.
Further investigation of badminton motion can assist us in designing tailored guidance for individuals with different skill levels.
Or the system's capability will extend by employing other stroke models for training.

\section{Conclusion}
In this paper, we have introduced a badminton backhand short service self-coaching system giving guidance and real-time feedback to the user.
We summarized the open kinetic chain characteristics of the backhand short service and conducted a study to quantify a service model, which is the basis of our training system.
We also evaluated the effectiveness and usability of the system through a user study.
The result shows that our system can help a novice user to fine-tune the exertion approach in a short-term practice, although we can not immediately promote the service performance.
From the study result, we specify several principles of designing a training system.
Eventually, we expect the system can include other strokes and reduce the motion capture cost to help more beginner to learn badminton in the future.

\begin{acks}
This work was supported by National Science and Technology Council in Taiwan (112-2636-E-002-002).
We extend our sincere appreciation to Ping-Yi Wang for her invaluable contributions during the early prototyping phase of the project, which played a crucial role in shaping the direction and success of our research.
We are also grateful to all the participants who graciously took part in the formative study, providing valuable data, insights, and feedback, which have been critical in advancing our understanding in this field.
Furthermore, we would like to express our special thanks to all the participants who joined the study under the challenging circumstances of the COVID-19 pandemic, as their engagement and contribution are greatly appreciated.
\end{acks}

\bibliographystyle{ACM-Reference-Format}
\bibliography{reference}

\appendix

\section{Kinetic Variables}
\label{app:kvdetail}
The following list is the name, description, and expression of variables that help us to derive the target kinetic variables.

\begin{itemize}
    \item \textbf{Racket major} \\ The direction from the bottom to the top of the racket \\ $$\vec{racket\ top} - \vec{racket\ bottom}$$ \\
    \item \textbf{Racket side} \\ The direction from the middle to the side of the racket head \\ $$\vec{racket\ side} - \vec{racket\ middle}$$ \\
    \item \textbf{Racket normal} \\ The normal vector of the racket head \\ $$\vec{racket_{major}} \times \vec{racket_{side}}$$ \\
    \item \textbf{Wrist} \\ The user's wrist of the hand holding the racket \\ $$\frac{\vec{WRA} + \vec{WRB}}{2}$$ \\
    \item \textbf{Forearm} \\ The vector from the elbow to the wrist \\ $$\vec{wrist} - \vec{elbow}$$ \\
    \item \textbf{Upper arm} \\ The vector from the shoulder to the elbow \\ $$\vec{elbow} - \vec{shoulder}$$ \\
    \item \textbf{Racket pitch angle} \\ The angle between the racket normal and the transverse plane \\ $$\arccos{90 - (\frac{Y_{racket\ normal}}{|\vec{racket\ normal}|})}$$ \\
    \item \textbf{Wrist angle} \\ The angle between the forearm and the racket major \\ $$\arccos{(\frac{\vec{forearm}\cdot\vec{racket_{major}}}{|\vec{forearm}||\vec{racket_{major}}|})}$$ \\
    \item \textbf{Elbow angle} \\ The angle between the upper arm and the forearm \\ $$\arccos{(\frac{\vec{forearm}\cdot\vec{-upper\ arm}}{|\vec{forearm}||\vec{-upper\ arm}|})}$$ \\
    \item \textbf{Shoulder angle} \\ The angle between the upper arm and $(0, -1, 0)$ \\ $$\arccos{(\frac{-Y_{upper\ arm}}{|\vec{upper\ arm}|})}$$ \\
\end{itemize}

\section{Time Variation of Kinetic Variables}
\label{app:kvvariation}
The following figures (Figure~\ref{fig:pitchraw} $\sim$ \ref{fig:shoulderraw}) are the changes in the racket pitch angle, racket height, racket speed, wrist angle, elbow angle, and shoulder angle during three training sessions.
The trials in the three sessions are colored red, green, and blue, respectively.
We determined the jitter service of all users according to the count of local extreme values (turning points) session by session, and we labeled all the jitters shots with a dashed line.
In racket pitch angle and speed, we examined the jitters only in the forward swing part because we care about the ending result of these two variables rather than their process.
Subsequently, we illustrated the regression of each session with a bold solid line.

\begin{figure}[!h]
    \centering
    \includegraphics[width=\linewidth]{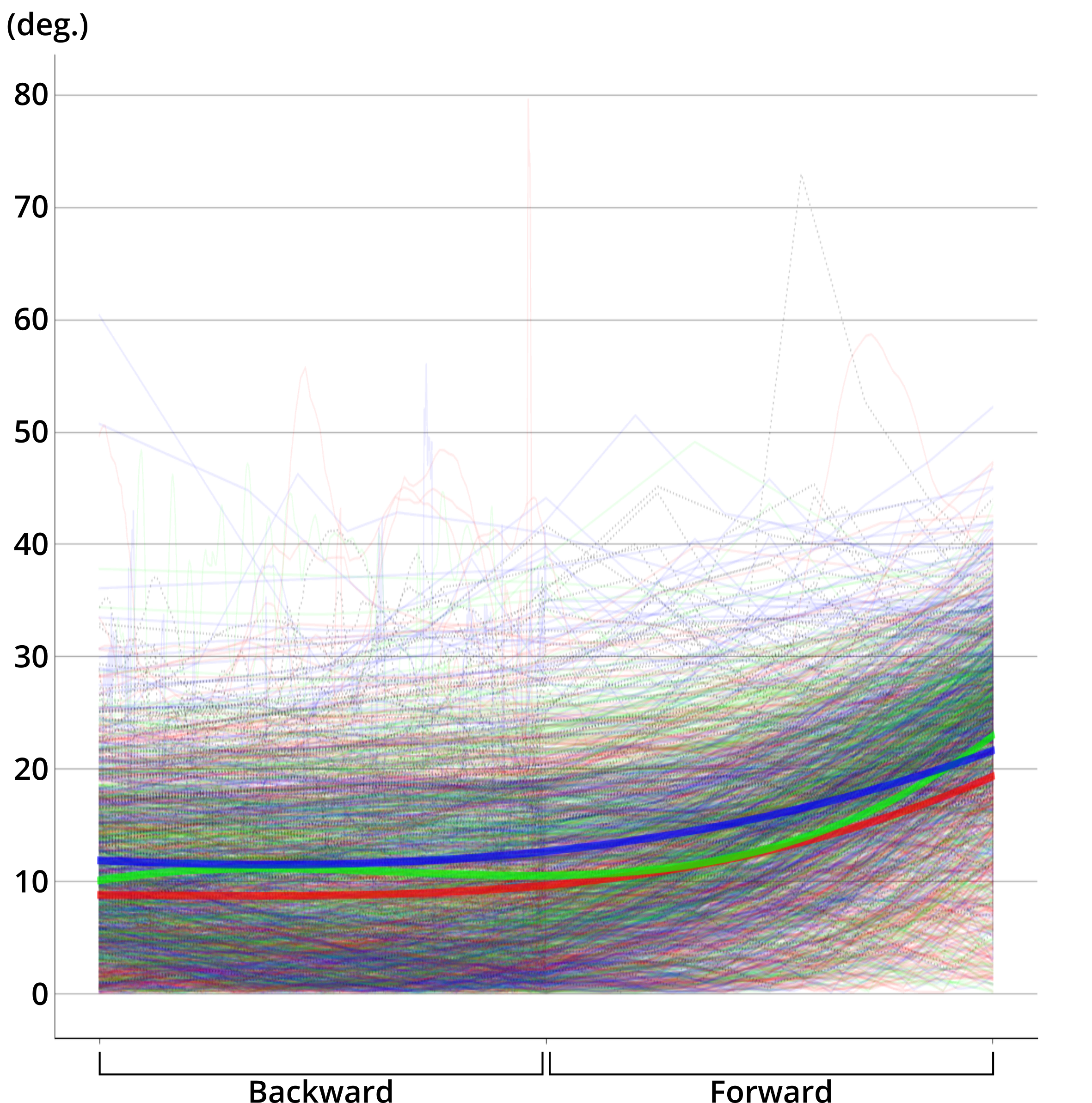}
    \caption{
    The change in the racket pitch angle.
    }
    \label{fig:pitchraw}
\end{figure}
\begin{figure}[!h]
    \centering
    \includegraphics[width=\linewidth]{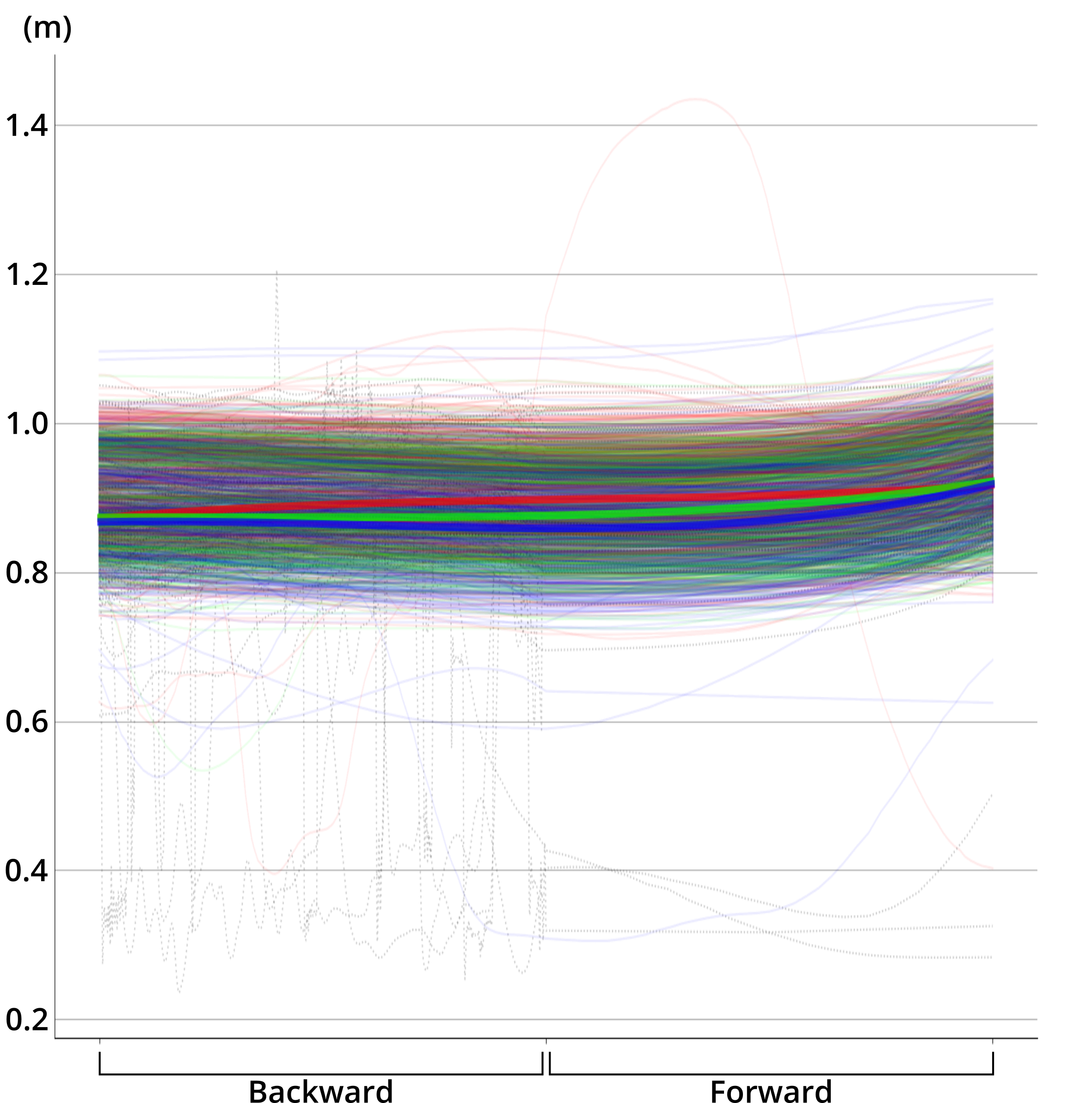}
    \caption{
    The change in the racket height.
    }
    \label{fig:heightraw}
\end{figure}
\begin{figure}[!h]
    \centering
    \includegraphics[width=\linewidth]{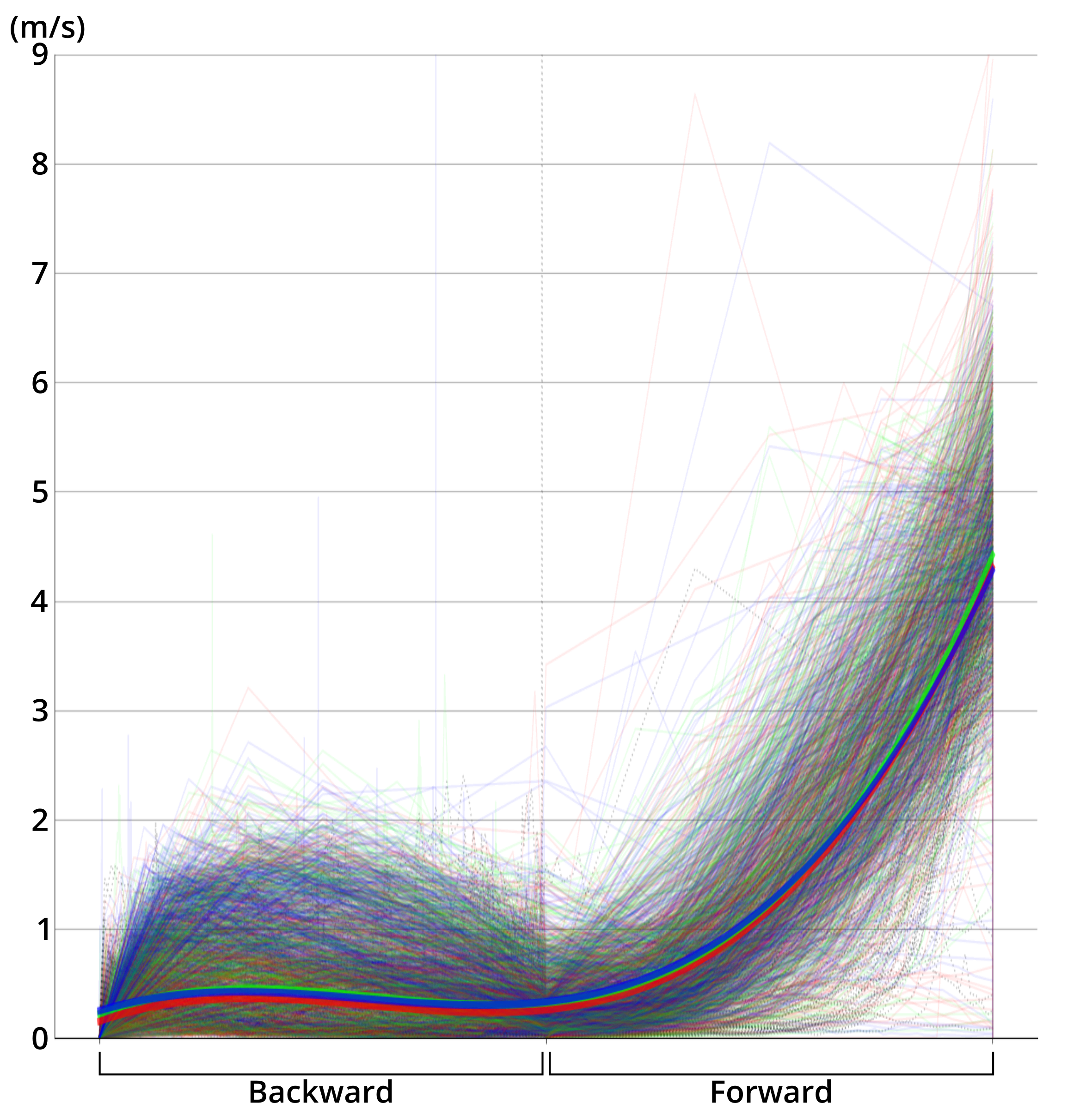}
    \caption{
    The change in the racket speed.
    }
    \label{fig:speedraw}
\end{figure}
\begin{figure}[!h]
    \centering
    \includegraphics[width=\linewidth]{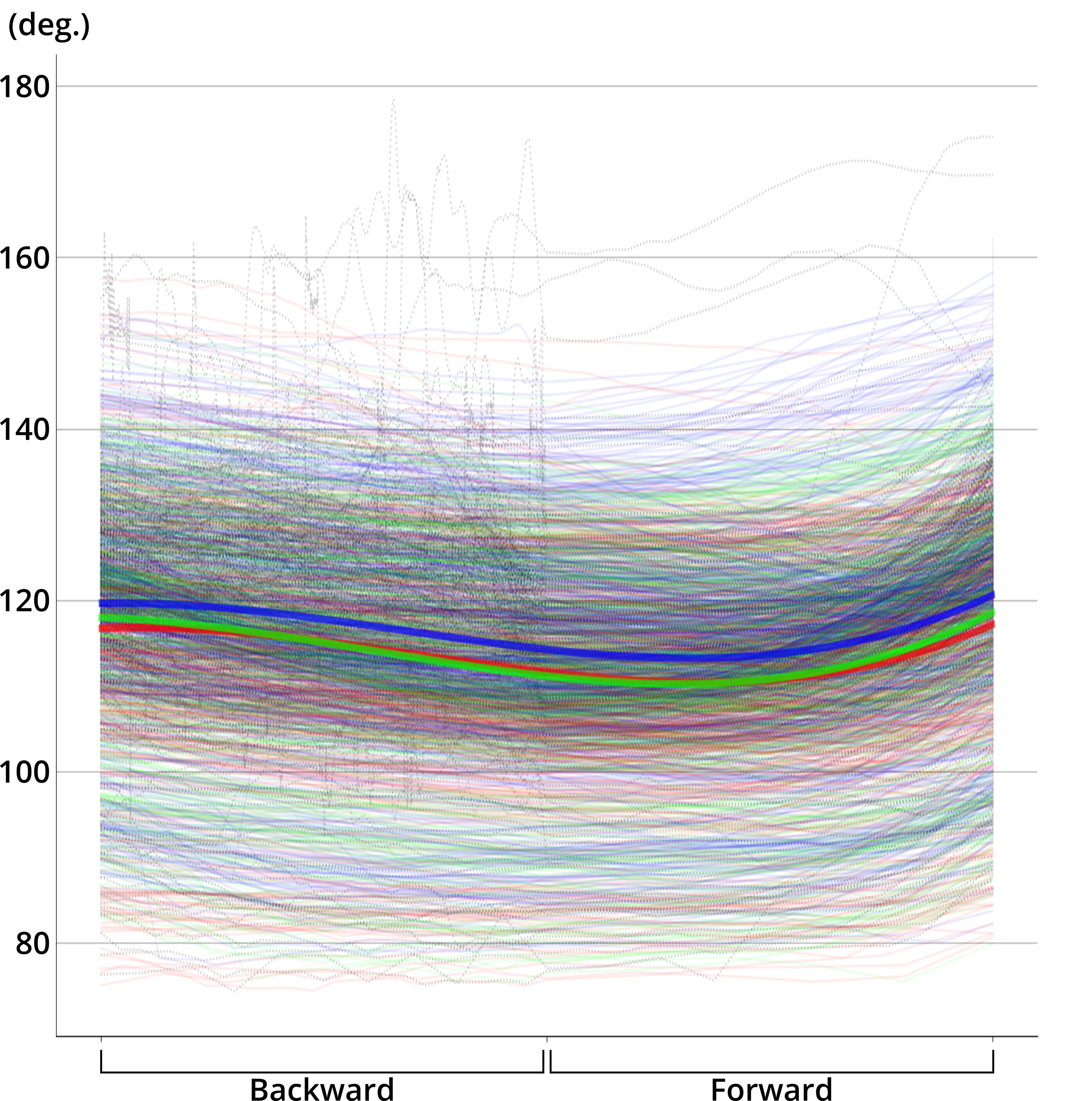}
    \caption{
    The change in the wrist angle.
    }
    \label{fig:wirstraw}
\end{figure}
\begin{figure}[!h]
    \centering
    \includegraphics[width=\linewidth]{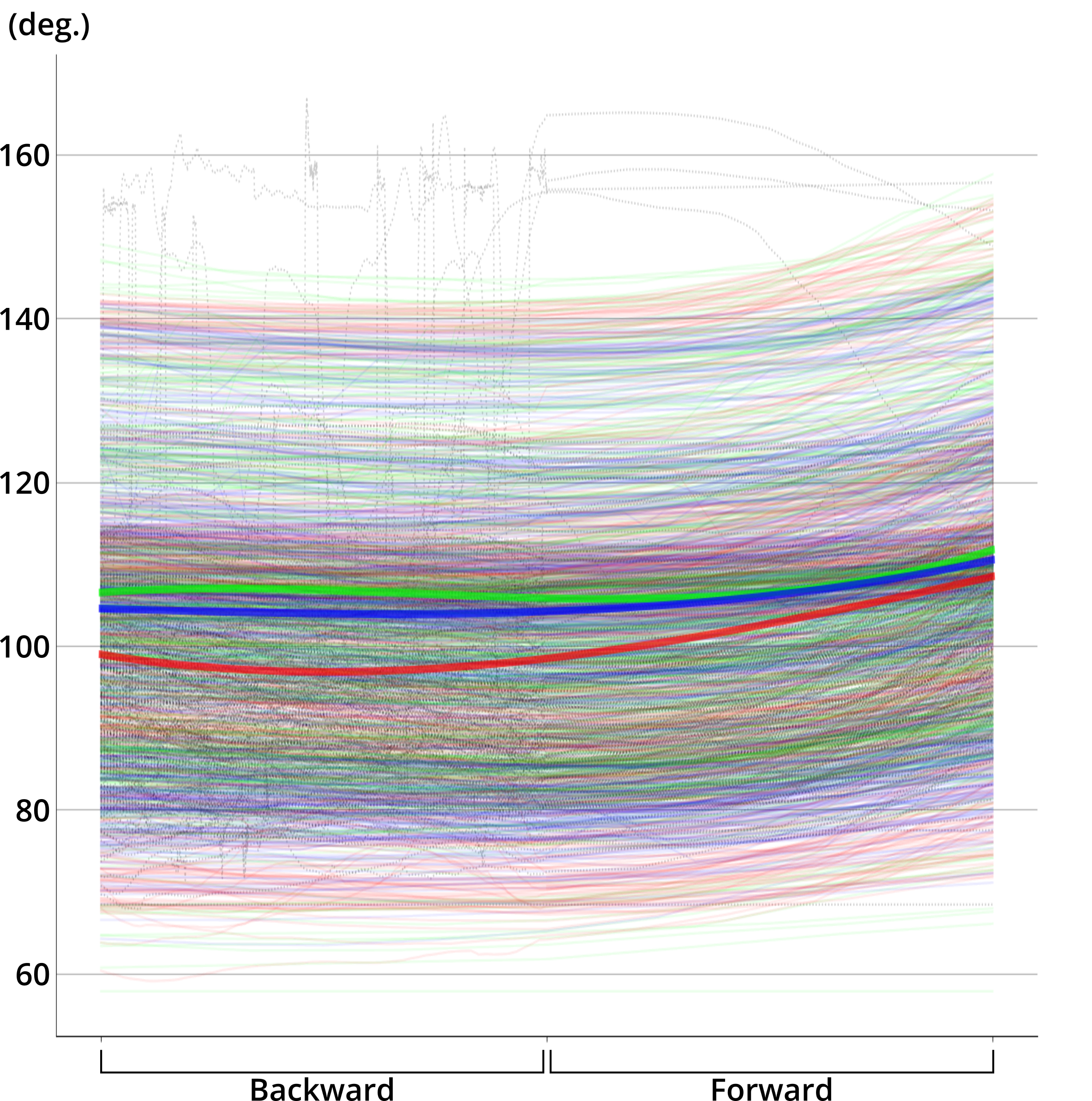}
    \caption{
    The change in the elbow angle.
    }
    \label{fig:elbowraw}
\end{figure}
\begin{figure}[!h]
    \centering
    \includegraphics[width=\linewidth]{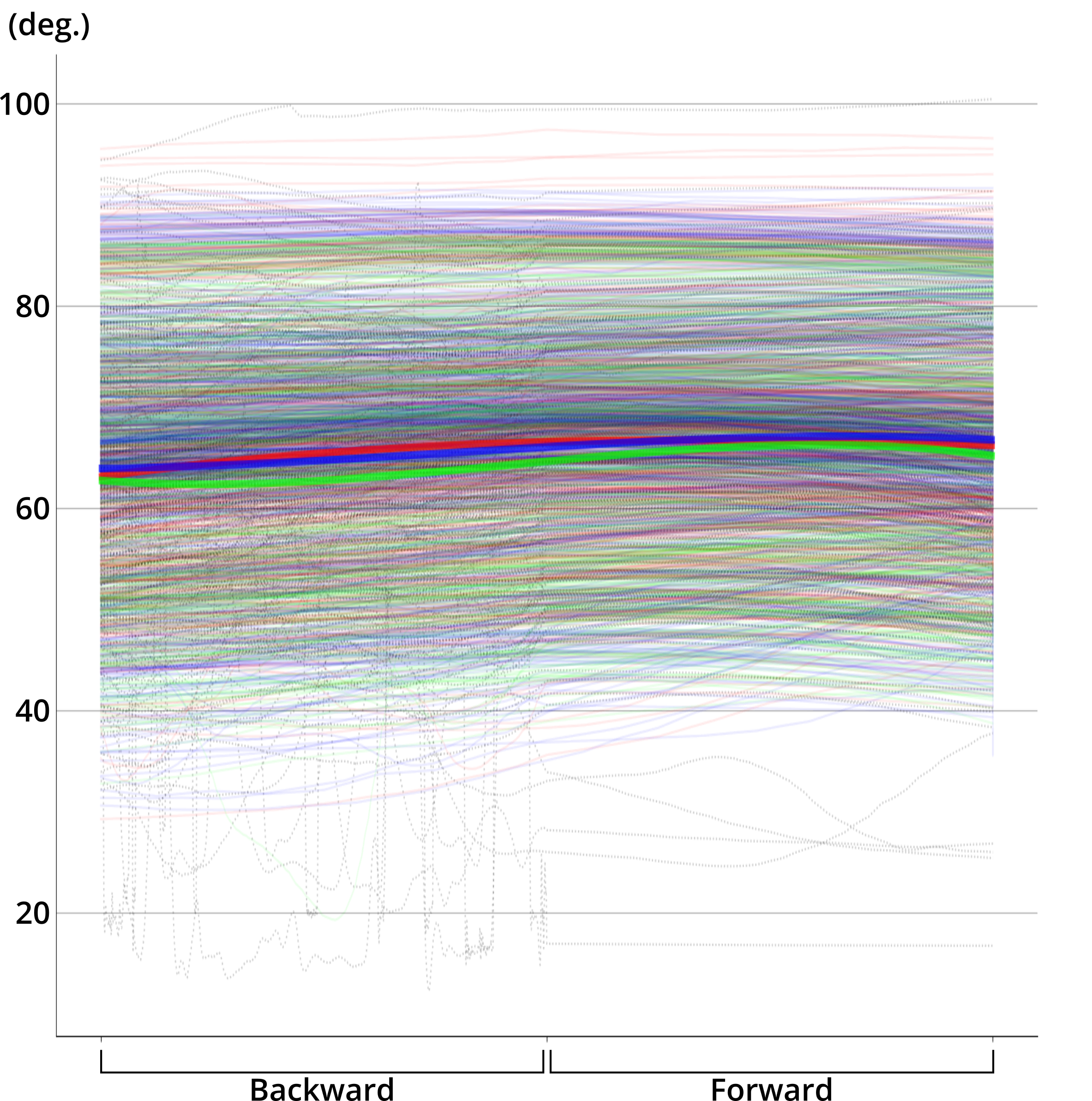}
    \caption{
    The change in the shoulder angle.
    }
    \label{fig:shoulderraw}
\end{figure}

\section{Racket speed in translation and rotation}
\label{app:racketspeed}
The racket speed results from the player's translational and rotary exertion.
Therefore, while the player swings with a constant force and exertion time, the racket speed should be between the speed in translational motion and rotary motion.

Consider a player swinging a racket weighted $m$ with a fixed force $F$ in time $T$ from a standstill.
The direction of $F$ is fixed in translation and always perpendicular to the racket's major axis in rotation.
When there is only translation, the racket speed is:
$$
v_{t} = a \cdot T,
$$
where $a$ is the linear acceleration and can be derived from Newton's second law of motion $F = ma$.
In contrast, when there is only rotation, we can calculate the racket speed from its angular speed $\omega$ at time $T$:
$$
v_{r} = \omega \cdot r_{top},
$$
where $r_{top}$ is the distance from the grip to the racket top.
Next, $\omega$ can be derived from the angular acceleration $\alpha$ by the definition $\alpha = \frac{\Delta\omega}{\Delta t}$.
Hence,
$$
v_{r} = \alpha \cdot T \cdot r_{top}.
$$
Note that $\alpha = \frac{\Delta \omega}{\Delta t} = \frac{\Delta v}{r \cdot \Delta t} = \frac{a}{r}$, where $r = r_{c}$ is the distance between the racket mass center and the grip here.
Now we can derive the speed in rotary motion:
$$
v_{r} = \frac{a}{r_{c}} \cdot T \cdot r_{top}.
$$
Substitute $a \cdot T$ with $v_{t}$:
$$
v_{r} = v_{t} \cdot \frac{r_{top}}{r_{c}}.
$$
Since $r_{top} \geq r_{c}$.
Thus,
$$
|v_{r}| \geq |v_{t}|.
$$

\end{document}